\documentclass[twocolumn,showpacs,aps,prb,preprintnumbers,amsmath,amssymb]{revtex4}
\usepackage{graphicx}% Include figure files
\usepackage{dcolumn}% Align table columns on decimal point
\usepackage{bm}% bold math
\usepackage{epsfig}

\begin{document}

%\preprint{APS/123-QED}

\title{
Boundary-induced violation of the Dirac fermion parity and its signatures \\   
in local and global tunneling spectra of graphene   
}

\author{Grigory Tkachov and Martina Hentschel}
\affiliation{
Max Planck Institute for the Physics of Complex Systems, Dresden, Germany}
%\date{\today}% It is always \today, today,
             %  but any date may be explicitly specified

\begin{abstract}
Extended defects in graphene, such as linear edges, break the translational invariance and  
can also have an impact on the symmetries specific to massless Dirac-like quasiparticles in this material.
The paper examines the consequences of a broken Dirac fermion parity 
in the framework of the effective boundary conditions varying from 
the Berry-Mondragon mass confinement to a zigzag edge.   
The parity breaking reflects the structural sublattice asymmetry of zigzag-type edges and 
is closely related to the previously predicted time-reversal symmetric edge states. 
We calculate the local and global densities of the edge states and show that they carry a specific polarization, 
resembling, to some extent, that of spin-polarized materials. 
The lack of the parity leads to a nonanalytical particle-hole asymmetry in the edge-state properties.  
We use our findings to interpret  recently observed tunneling spectra in zigzag-terminated graphene.  
We also propose a graphene-based tunneling device where the particle-hole asymmetric edge states result in 
a strongly nonlinear conductance-voltage characteristics, 
which could be used to manipulate the tunneling transport.  
\end{abstract}


\maketitle

\section{Introduction}
\label{Intro}
 
In condensed matter systems with nodal fermionic spectra, 
quantum description of low-energy excitations can resemble that of the ultrarelativistic electron.
The crystal space group imposes a generic restriction on such quasiparticles known as fermion doubling:  
they come in pairs of opposite chirality species that can be mapped to 
the conventional "right-handed" (RH) and "left-handed" (LH) fermions of the Dirac theory.~\cite{NN81}
The most recently studied examples are graphene, 
where two distinct Fermi points in the Brillouin zone
give rise to both chiral species,~\cite{Semenoff84,Novoselov05} 
and 2D HgTe quantum wells where spin-orbit coupling
effectively results in a pair of the RH and LH fermions at low energies.~\cite{Koenig07,Koenig08,Schmidt09}
The fermion doubling brings the symmetry with respect to the exchange of the chiralities, RH $\leftrightarrows$ LH, 
related to the parity symmetry of the Dirac equation.~\cite{Itzykson} 
It is of considerable interest to investigate the consequences of the violation of such a symmetry, 
since they could be observable in materials where quasiparticles imitate Dirac electrons. 
Besides, the Dirac fermion parity is distinct from other discrete symmetries 
(e.g. time-reversal invariance), and, therefore, through its violation 
one could achieve additional control over electronic properties of the material.   

It has been noticed~\cite{McCann04,Akhmerov08} that discrete symmetries 
of a Dirac fermion system can be broken, along with the translational invariance, 
by the boundaries of the system. 
In this paper we focus on the parity violation due to such a boundary effect 
and suggest how to detect and, possibly, use it in electronic devices.        

Our main assumption is that the boundary does not cause scattering 
between the opposite chirality quasiparticles. 
To model this we use effective boundary conditions,~\cite{McCann04,Akhmerov08}   
interpolating between the infinite mass confinement~\cite{Berry87} 
and the zigzag graphene boundary.~\cite{Fujita96,Armchair} 
The parity breaking occurs as long as the boundary deviates from the infinite mass confinement toward  
the zigzag edge. This is due to the structural sublattice asymmetry: 
a zigzag-type crystal face has unequal numbers of sites from 
the two sublattices of the honeycomb structure.~\cite{Fujita96,Akhmerov08}  
More generally, the parity in this continuum model can cannot be preserved simultaneously 
with the time-reversal invariance near the edge of the system. 
This is closely related to the existence of time-reversal symmetric, 
propagating edge states~\cite{Fujita96}   
which have become a topic of vigorous graphene-related research 
(see, e.g., Refs.~\onlinecite{Waka00,Koba05,Niimi06,Peres06,Sasaki06,Brey06,Rycerz07,GT07},\onlinecite{Akhmerov08,Wimmer08}).
The time-reversal symmetry requires that the edge states from the different valleys 
propagate in the opposite directions, forming a Kramers pair at a given energy.
As a result, even in the absence of the intervalley scattering 
the problem does not reduce to a single valley, in the sense 
that the valley contributions to observables are not identical. 
For the low-energy states (imitating the RH and LH fermions), 
this implies broken Dirac fermion parity.

We intend to demonstrate several new properties of the broken-parity edge states: 

{\bf (i)} nonanalytic particle-hole asymmetry of local and global densities of states,

{\bf (ii)} time-reversal invariant pseudospin polarization 
(which in graphene is associated with the sublattice degree of freedom), and

{\bf (iii)} asymmetric nonlinear bias-voltage dependence of the tunneling conductance.
\newline{
In view of the progress in experimental control over graphene edges,~\cite{Girit09} 
this material is particularly suitable to test our findings. 
Below we discuss in more detail the connection between our results 
and the ongoing graphene-related research. 
}

Our finding {\bf (i)} can be tested by means of scanning tunneling spectroscopy (STS) of the density of states (DOS).  
In fact, some STS experiments~\cite{Koba05,Niimi06} have already  
reported a particle-hole asymmetric DOS with a peak at -20-50 meV  for monoatomic zigzag graphene edges. 
By contrast Li et al.~\cite{Li08} have observed a symmetric linear DOS in graphene bulk. 
In our model, the crossover from an asymmetric edge DOS to a symmetric bulk one follows naturally 
from the existence of the broken-parity Dirac fermion edge states. 
The results of experiments~\cite{Koba05,Niimi06} could therefore provide 
some evidence for the Dirac fermion parity violation. 
Such an interpretation is also supported by the observation 
that the position of the DOS peak in the experiment of Niimi et al.~\cite{Niimi06}   
can be described very accurately by our model. 
This is achieved by taking into account not only the structural asymmetry 
but also a potential-energy difference between the sublattices~\cite{Akhmerov08} 
that generates weakly dispersive edge states rather than the singular zero-energy band.~\cite{Fujita96}  
It was shown earlier~\cite{Sasaki06} that the next-nearest-neighbor hopping on the honeycomb lattice 
could also result in a particle-hole asymmetric edge DOS.   
However, this symmetry-breaking mechanism results in the DOS peak at significantly larger energies, 
of the order of next-nearest-neighbor hopping energy $\approx 300$ meV.

Another our result {\bf (ii)} demonstrates that broken-parity edge states carry 
a time-reversal-invariant pseudospin polarization. 
This agrees with the general perception that   
in graphene electronic properties and those arising from 
the sublattice degree of freedom (pseudospin) are interrelated. 
However, there is a great deal of uncertainty as to how such a relation can be studied. 
The specific feature of our edge problem is that it is possible to establish 
a one-to-one correspondence between the pseudospin polarization and the edge DOS.  
We suggest that the pseudospin polarization can be detected via measurements 
of the electric conductance in lateral tunnel contacts between zigzag-terminated graphene 
and a suitably chosen metallic electrode.
 
Our proposal is based on finding {\bf (iii)} that  
the edge-state (i.e. polarization-dependent) contribution to the conductance is  
asymmetric with respect to the bias voltage. 
Therefore, it can be separated from the symmetric contribution of the bulk graphene states.   
In addition,  the edge-state tunnel conductance turns out to be strongly nonlinear: 
It exhibits kink-like switching as the sign of the voltage reverses.  
Such a behavior could serve as a prototype for the potentially useful electronic functionality. 

The outline of the paper is as follows. 
In Sec.~\ref{Model} we formulate the boundary problem for the Green's function 
of the Dirac equation and discuss the role of the parity symmetry. 
In Sec.~\ref{Local} we analyze the local DOS and pseudospin polarization, 
and compare our results for the local DOS with the experimental data of Niimi et al.~\cite{Niimi06}
Section~\ref{Global} describes the relation between the global edge DOS and 
pseudospin polarization and their tunneling spectroscopy.  
The last section~\ref{Discussion} summarizes our results and  
discusses their validity as well as possible applications.

\section{Dirac fermions in 2D semi-space: Broken parity and edge states}
\label{Model}

Edge-state spectroscopy usually deals with isolated edges in large samples 
where finite-size effects are presumably irrelevant.~\cite{Niimi06,Koba05,Girit09} 
We model this by considering a boundary problem 
for a Dirac fermion retarded Green's function in a 2D semi-space 
$-\infty < x <\infty$, $0<y<\infty$:  
\begin{eqnarray}
&
( \varepsilon I - v\gamma^5\mbox{\boldmath$\Sigma$}\mbox{\boldmath$p$})
G_{\varepsilon}( {\bf r},{\bf r}^\prime )=
\delta( {\bf r}-{\bf r}^\prime ), 
&
\label{Eq}
\end{eqnarray}
\begin{eqnarray}
&
G_{\varepsilon}=
\left( \frac{ I +  \gamma^5 }{ 2 }\mbox{\boldmath$\Sigma$}{\bf n}_+  + 
       \frac{ I -  \gamma^5 }{ 2 }\mbox{\boldmath$\Sigma$}{\bf n}_- 
\right)
G_{\varepsilon}\big|_{y=0},\,\,  {\bf n}^2_\pm =1,
&
\label{BC}
\end{eqnarray}
with $G|_{y\to\infty}$ being finite. 
In Eq.~(\ref{Eq}) $\mbox{\boldmath$p$}=-i\hbar(\partial_x, \partial_y, 0)$, 
$\varepsilon$ and $v$ are the 2D momentum operator, energy and velocity near a Fermi point.  
In view of the further analysis of the parity symmetry, equations~(\ref{Eq}) and (\ref{BC}) are both expressed 
in terms of the chirality, $\gamma^5$ and effective spin, $\mbox{\boldmath$\Sigma$}$    
through the Dirac matrices:~\cite{Itzykson} 
\begin{eqnarray}
&
\gamma^5=i\gamma^0\gamma^1\gamma^2\gamma^3=\tau^3\otimes\sigma^0, 
\quad
\mbox{\boldmath$\Sigma$}=\gamma^5\gamma^0\mbox{\boldmath$\gamma$}=\tau^0\otimes\mbox{\boldmath$\sigma$},
&
\label{Spin}\\
&
\gamma^0=-\tau^1\otimes\sigma^0,\quad
\mbox{\boldmath$\gamma$}=i\tau^2\otimes\mbox{\boldmath$\sigma$}.
&
\label{Gamma}
\end{eqnarray}
We introduce the two sets of Pauli matrices, $\sigma^{1,2,3}$ and $\tau^{1,2,3}$, 
and the corresponding unit matrices, $\sigma^0,\tau^0$ and $I=\tau^0\otimes\sigma^0$. 
In graphene, $\sigma^{1,2,3}$ represent the two sublattices of the honeycomb structure, 
while $\tau^{1,2,3}$ act in the valley space.    
The eigenstates and eigenvalues ($\tau=\pm 1$) of the hermitean matrix 
$\gamma^5$ conventionally define the right-handed (RH, $+$) and 
left-handed (LH, $-$) quasiparticles~\cite{Itzykson}. 
In Eqs.~(\ref{Eq}) and (\ref{BC}) they are described by the projected Green's functions 
$\frac{ 1 }{ 2 }(I \pm  \gamma^5)G_{\varepsilon}$.

The boundary condition, Eq.~(\ref{BC}) 
ensures vanishing of the particle current across the edge~\cite{McCann04,Akhmerov08} 
(i.e. no Klein tunneling~\cite{Katsnelson06}). 
It is diagonal in chirality space with the RH and LH blocks parametrized 
by three-dimensional unit vectors, ${\bf n}_\tau={\bf n}_{\pm}$, 
orthogonal to the boundary normal.~\cite{McCann04,Akhmerov08}
In graphene, where $\tau=\pm 1$ label the valleys,  
the case of the zigzag edge corresponds to~\cite{Akhmerov08} 
\begin{eqnarray}
{\bf n}_{+}=-{\bf n}_{-}={\bf\hat z}, 
\quad
G_{\varepsilon}=\gamma^5 \Sigma^3G_{\varepsilon}\big|_{y=0},
\label{Z}
\end{eqnarray}
where ${\bf\hat z}$ is the out-of-plane unit vector. 
This implies that one of the sublattice Green's functions must vanish at the edge, 
reflecting the structural sublattice asymmetry of the zigzag boundary.~\cite{Fujita96}
The other nontrivial limit is 
\begin{eqnarray}
{\bf n}_{+}={\bf n}_{-}={\bf\hat x},
\quad 
G_{\varepsilon}=\Sigma^1G_{\varepsilon}\big|_{y=0}.
\label{MC}
\end{eqnarray}
It is the infinite mass confinement of Berry and Mondragon~\cite{Berry87} 
(${\bf\hat x}$ is the unit vector along the edge). 
As shown in Ref.~\onlinecite{Akhmerov08},
the intermediate case, when ${\bf n}_{\pm}$ interpolate between (\ref{Z}) and (\ref{MC}), 
can be treated as a zigzag edge where the {\em structural} sublattice asymmetry 
coexists with a {\em potential} (energy) sublattice asymmetry   
that could result from electron-electron interactions 
within atomic distances near the edge.~\cite{Son06}  
This does not however exhaust the applicability of the boundary condition (\ref{BC}), 
since it can be derived from the only requirement that the Dirac particle current is zero 
in the normal direction at the edge.~\cite{McCann04,Akhmerov08} 

Unless restricted to the infinite mass confinement case (\ref{MC}), 
the boundary parameters are not identical, ${\bf n}_+\not ={\bf n}_-$, 
which is verified below on the basis of the time-reversal ($\cal T$) symmetry [see, Eq.~(\ref{K})].  
Therefore, boundary condition~(\ref{BC}) 
(and, in particular, Eq.~(\ref{Z})) explicitly contains the chirality, $\gamma^5$.
This violates the symmetry under the exchange of the RH and LH quasiparticles,
\begin{equation} 
G_{\varepsilon}\to \gamma^0G_{\varepsilon}\gamma^0, 
\label{P1}
\end{equation}
since it reverses the sign of $\gamma^5$ 
(In contrast, $\mbox{\boldmath$\Sigma$}$ is even under such operation). 
On the other hand, the RH $\leftrightarrows$ LH exchange 
is involved in the parity transformation,~\cite{Itzykson} 
\begin{eqnarray} 
G_{\varepsilon}({\bf r},{\bf r}^\prime)\to 
\gamma^0 G_{\varepsilon}(-{\bf r},-{\bf r}^\prime) \gamma^0,
\label{P2}
\end{eqnarray}
and in the particle-hole conjugation, 
\begin{eqnarray} 
G_{\varepsilon}({\bf r},{\bf r}^\prime)
\to -\gamma^0 G_{-\varepsilon}({\bf r},{\bf r}^\prime) \gamma^0, 
\label{P3}
\end{eqnarray}
both leaving the Dirac equation (\ref{Eq}) invariant.
Therefore, if  boundary condition (\ref{BC}) deviates from the infinite mass confinement (\ref{MC}),  
our boundary problem exhibits no parity invariance and, in view of Eq.~(\ref{P3}), 
no particle-hole symmetry. The symmetry breaking persists in the limit of the zigzag edge, Eq.~(\ref{Z}). 
We therefore conclude that the parity breaking is due to 
the structural sublattice asymmetry of the zigzag-type lattice termination.~\cite{Fujita96} 
For practical calculations, we need to take into account deviations of ${\bf n}_\pm$ from ${\bf\hat z}$
(e.g. due to potential-energy sublattice asymmetry~\cite{Akhmerov08}),
because it eliminates the Green's function singularity at $\varepsilon=0$ 
characteristic to dispersionless zero-energy edge states. 

The connection between the parity breaking and 
the existence of the edge states  
can be established by explicit calculation of the Green's function from Eqs.~(\ref{Eq}) and (\ref{BC}). 
The calculation details are given elsewhere.~\cite{GT07}  
Here we present the final result:       
\begin{eqnarray}
&&
G_{\varepsilon}({\bf r},{\bf r}^\prime)=\sum_{\tau=\pm 1, k}
\left( \frac{I+\tau\gamma^5}{2} \right)
\left( I+\frac{\tau v}{\varepsilon}\mbox{\boldmath$\Sigma$}
\mbox{\boldmath$p$} \right)
\nonumber\\
&&
\times\left( G^{(0)}_{\varepsilon\tau k}(y,y^\prime)I + 
G^{(3)}_{\varepsilon\tau k}(y,y^\prime)\Sigma^3 
\right)
\frac{{\rm e}^{ik(x-x^\prime)} }{L},
\label{G}
\end{eqnarray}
\begin{eqnarray}
&&
G^{(0)}_{\varepsilon\tau k}(y,y^\prime)=\frac{\varepsilon}{2\hbar^2v^2q}
\left(
{\rm e}^{-q(y+y^\prime)} - {\rm e}^{ -q|y-y^\prime| }
\right)
\nonumber\\
&&
+
\frac{ q + k n_{z\tau} }{ 2( \varepsilon  - \hbar v \tau k n_{x\tau} ) }
\,{\rm e}^{-q(y+y^\prime)},
\label{G0}
\end{eqnarray}
\begin{eqnarray}
&&
G^{(3)}_{\varepsilon\tau k}(y,y^\prime)=
\frac{k + q n_{z \tau} -\tau\varepsilon n_{x \tau}/\hbar v  }
{2( \varepsilon  - \hbar v \tau k n_{x \tau} )}
\,{\rm e}^{-q(y+y^\prime)},
\label{G3}
\end{eqnarray} 
where $q=\sqrt{ k^2-\varepsilon^2/\hbar^2 v^2 }$ and $k$ is the wave number. 
In Eqs.~(\ref{G0}) and (\ref{G3}) the edge states are described by the terms 
with the pole at $\varepsilon=\hbar v \tau k n_{x\tau}$. 
Let us examine, for instance, Eq.~(\ref{G3})  near the pole: 
\begin{equation}
G^{(3)}_{\varepsilon \tau k}(y,y^\prime)\approx 
-\frac{n_{z\tau}\Theta( kn_{z\tau} )}{\varepsilon  - \hbar v \tau k n_{x\tau}} 
\partial_y\,{\rm e}^{-|kn_{z\tau}|(y+y^\prime)}.
\label{G3_1}
\end{equation} 
Clearly, the pole exists only if the unit step function $\Theta(kn_{z\tau})$ 
is nonzero, which determines the spectrum as 
\begin{equation}
\varepsilon_{\tau k}=\hbar v \tau k n_{x\tau}, \quad\quad kn_{z\tau}>0. 
\label{Spectrum}
\end{equation} 
These equations are not yet restricted by the ${\cal T}$ symmetry. 
The ${\cal T}$-symmetric spectrum follows from the condition that 
both equations (\ref{Spectrum}) are invariant under the simultaneous reversal of 
the chirality and wave-vector,  $\tau, k \to -\tau, -k$. 
This imposes the following restrictions on ${\bf n}_\tau$: 
\begin{eqnarray}
n_{x\tau}=n_x, \quad n_{z\tau}=\tau n_z, 
\quad {\bf n}=(n_x, 0, n_z),
\,\, {\bf n}^2=1,
\label{K}
\end{eqnarray} 
leaving a single free boundary parameter - 
the direction of the unit vector ${\bf n}$. 
The edge-state spectrum is now manifestly Kramers degenerate and 
particle-hole asymmetric:~\cite{Akhmerov08}
 \begin{equation}
\varepsilon_{\tau k}=\hbar v \tau k n_x, \quad\quad \tau kn_z>0. 
\label{Spectrum1}
\end{equation} 
The role of the parity breaking is quite apparent from the behavior of 
the edge-state Green's function~(\ref{G3_1}): 
it vanishes identically for the infinite mass confinement ($n_z=0$) 
which preserves the parity symmetry (see, also Eq.~(\ref{MC})).     

The knowledge of the spectrum (\ref{Spectrum1}) is not sufficient 
to interpret the STS measurements,~\cite{Koba05, Niimi06} 
as they provide information on the local DOS rather than dispersion $\varepsilon_{\tau k}$. 
In the next section we use the full Green's function (\ref{G}) 
to calculate the local DOS of the system. 
We will see that in addition to the exponentially localized states (\ref{G3_1})
there is another type of edge states decaying algebraically 
as a consequence of the lack of the energy gap in the 2D bulk. 
This distinguishes our system from, e.g., topological insulators 
where bulk excitations are fully gapped.~\cite{Kane05,Koenig07,Koenig08}

\section{Local DOS and pseudospin polarization}
\label{Local}

\subsection{Particle-hole symmetry and the role of parity}
\label{Parity}

The spectral and pseudospin properties of the system are interrelated. 
Let us define the local DOS
\begin{equation}
\nu_\pm(\varepsilon,{\bf r})=-\frac{1}{2\pi}{\rm Im Tr}\,
(I \pm \gamma^5)G_{\varepsilon}({\bf r},{\bf r}), 
\label{nu_pm_def}
\end{equation} 
and local pseudospin polarizations
\begin{equation}
p_\pm(\varepsilon,{\bf r})=-\frac{1}{2\pi}{\rm Im Tr}\,
(I \pm \gamma^5)\Sigma^3 G_{\varepsilon}({\bf r},{\bf r})
\label{p_pm_def}
\end{equation} 
in terms of the RH and LH projections of the Green's function, $G_\varepsilon$. 
Tunneling spectra are determined by the total local DOS related to $G^{(0)}$ in Eq.~(\ref{G0}): 
\begin{eqnarray}
&
\nu(\varepsilon,{\bf r})=\nu_+ + \nu_-=
-\frac{2}{\pi L}\sum\limits_{\tau=\pm 1,k}
{\rm Im}G^{(0)}_{\varepsilon\tau k}(y,y).
&
\label{nu_def}
\end{eqnarray} 
Likewise, $p_+ + p_-=-\frac{1}{\pi}{\rm Im Tr}\,\Sigma^3 G_{\varepsilon}$ is the net pseudospin polarization. 
It vanishes by ${\cal T}$ symmetry, since Eqs.~(\ref{K}) yield $p_-= - p_+$. 
As a $\cal T$-invariant  characteristic of the pseudospin properties, 
we use the chiral pseudospin polarization (CPP) related to  $G^{(3)}$ in Eq.~(\ref{G3}):
\begin{eqnarray}
&
p(\varepsilon,{\bf r})=p_+-p_-=-\frac{1}{\pi}{\rm Im Tr}\,\gamma^5\Sigma^3 G_{\varepsilon}({\bf r},{\bf r})=
&
\nonumber\\
&
=-\frac{2}{\pi L}\sum\limits_{\tau=\pm 1,k}\tau {\rm Im}G^{(3)}_{\varepsilon\tau k}(y,y).
&
\label{p_def}
\end{eqnarray} 
Integrating over $k$ in Eqs.~(\ref{nu_def}) and (\ref{p_def}), 
we obtain  
\begin{eqnarray}
&&
\nu(\varepsilon,y)=\frac{ 2|\varepsilon| }{ \pi \hbar^2 v^2 } -
	\sum_{\tau=\pm 1} \frac{ \Theta\left( \varepsilon\, \tau n_{x\tau} n_{z\tau} \right) 
        }{h v |n_{x\tau}| }
        \partial_y
        {\rm e}^{
        -\frac{ 2y }{ \hbar v } \left|\varepsilon \frac{ n_{z\tau}  }{ n_{x\tau}  } \right|  
        }
	\nonumber\\
&&
        -\frac{ |\varepsilon|}{ \pi^2\hbar^2v^2 }
        \sum_{\tau=\pm 1} 
	\int_0^{ \frac{\pi}{2} } d\gamma\times  
        \label{nu}\\
&&
        \times\frac{
               n^2_{z\tau}\cos\left( \frac{2\varepsilon y}{\hbar v } \sin\gamma \right)
               +\tau n_{x\tau}n_{z\tau}\sin\gamma\sin\left( \frac{2\varepsilon y}{\hbar v } \sin\gamma \right)
	       }
	      { n^2_{z\tau}+\tan^2\gamma },
\nonumber	
\end{eqnarray}
\begin{eqnarray}
&&
	p(\varepsilon,y)=-
	\sum_{\tau=\pm 1} 
        \frac{ \tau n_{z\tau}\Theta\left( \varepsilon\, \tau n_{x\tau} n_{z\tau} \right) 
        }{h v |n_{x\tau}| }
        \partial_y {\rm e}^{
        -\frac{ 2y }{\hbar v} \left|\varepsilon \frac{  n_{z\tau} }{ n_{x\tau} } \right| 
        }
	\nonumber\\
	&&
        +\frac{|\varepsilon|}{\pi^2\hbar^2v^2}
        \sum_{\tau=\pm 1} 
	\int_0^{ \frac{\pi}{2}  }d\gamma \tan^2\gamma 
	\times 
        \label{p}\\
	&&
        \frac{     \tau n_{z\tau}\cos\left( \frac{2\varepsilon y}{\hbar v } \sin\gamma \right) 
                  +n_{x\tau}\sin\gamma\sin\left( \frac{2\varepsilon y}{\hbar v } \sin\gamma \right) }
	                   { n^2_{z\tau}+\tan^2\gamma }.
	\nonumber
\end{eqnarray}

It is now easy to see that the particle-hole symmetry is controlled by the parity.
In the broken-parity state with ${\bf n}_+\not={\bf n}_-$ given by Eqs.~(\ref{K}), 
the summation over chiralities $\tau=\pm 1$ in Eq.~(\ref{nu}) yields 
an asymmetric DOS as a function of energy, $\varepsilon$:

\begin{eqnarray}
&&
\nu(\varepsilon,y)=\frac{ 2|\varepsilon| }{ \pi \hbar^2 v^2 }-
	 \frac{ 2\Theta\left( \varepsilon\, n_{x} n_{z} \right) 
        }{h v |n_{x}| }
        \partial_y
        {\rm e}^{
 -\frac{ 2y }{ \hbar v } \left|\varepsilon \frac{ n_{z}  }{ n_{x}  } \right|  
        }
	\nonumber\\
&&
        -\frac{ 2|\varepsilon|}{ \pi^2\hbar^2v^2 } 
	       \int_0^{ \frac{\pi}{2} } d\gamma\times  
         \label{nu_1}\\
&&
        \times\frac{
               n^2_{z}\cos\left( \frac{2\varepsilon y}{\hbar v } \sin\gamma \right)
               +n_{x}n_{z}\sin\gamma\sin\left( \frac{2\varepsilon y}{\hbar v } \sin\gamma \right)
	       }
	      { n^2_{z}+\tan^2\gamma },
\nonumber	
\end{eqnarray}
The same is true for the CPP (\ref{p}): 
\begin{eqnarray}
&&
	p(\varepsilon,y)=- 
   \frac{ 2 n_{z}\Theta\left( \varepsilon\, n_{x} n_{z} \right) 
        }{h v |n_{x}| }
        \partial_y {\rm e}^{
        -\frac{ 2y }{\hbar v} \left|\varepsilon \frac{  n_{z} }{ n_{x} } \right| 
        }
	\nonumber\\
	&&
        +\frac{2|\varepsilon|}{\pi^2\hbar^2v^2} 
	\int_0^{ \frac{\pi}{2}  }d\gamma \tan^2\gamma 
	\times 
        \label{p_1}\\
	&&
        \frac{     n_{z}\cos\left( \frac{2\varepsilon y}{\hbar v } \sin\gamma \right) 
                  +n_{x}\sin\gamma\sin\left( \frac{2\varepsilon y}{\hbar v } \sin\gamma \right) }
	                   { n^2_{z}+\tan^2\gamma },
	\nonumber
\end{eqnarray}
For comparison, if the parity is preserved for ${\bf n}_+={\bf n}_-$,  
the DOS (\ref{nu}) appears to be an even function of energy:  

\begin{eqnarray}
&&
\nu(\varepsilon,y)=\frac{ 2|\varepsilon| }{ \pi \hbar^2 v^2 }-
	 \frac{ 1 }{ h v |n_{x}| }
        \partial_y
        {\rm e}^{ 
 -\frac{ 2y }{ \hbar v } \left|\varepsilon \frac{ n_{z}  }{ n_{x}  } \right|  
        }
	\nonumber\\
&&
        -\frac{ 2|\varepsilon|n^2_{z}}{ \pi^2\hbar^2v^2 } 
	       \int_0^{ \frac{\pi}{2} } d\gamma
	       \frac{
     \cos\left( \frac{2\varepsilon y}{\hbar v } \sin\gamma \right)
                   }
	                 { n^2_{z}+\tan^2\gamma }, 
\label{nu_2}	
\end{eqnarray}
However, the requirements for the parity symmetry, ${\bf n}_+={\bf n}_-$, 
are incompatible with conditions~(\ref{K}) for the ${\cal T}$ symmetry. 
The only exception is the infinite mass confinement limit $n_z\to 0$.  
In literature,~\cite{Fujita96,Son06,Wimmer08,Yazyev08} 
${\cal T}$-symmetry breaking on zigzag graphene edges has been discussed 
in connection with their possible intrinsic magnetism.  
It is still unclear whether the ${\cal T}$-symmetry breaking in the boundary condition, Eq.~(\ref{BC})
has anything to do with the edge magnetism.  
We will therefore limit our analysis to the ${\cal T}$-symmetric case (\ref{K}).

%%%%%%%%%%%%%%%%%%%%%%%%%%%%%%%%%%%%%%%%%%%%%%%%%%%%%%%%%%%%%%%%%%%%%
%%%%%%%%%%%%%%%%% LDOS %%%%%%%%%%%%%%%%%%%%%%%%%%%%%%%%%%%%%%%%%%%%%%
%%%%%%%%%%%%%%%%%%%%%%%%%%%%%%%%%%%%%%%%%%%%%%%%%%%%%%%%%%%%%%%%%%%%%
\begin{figure}[t]
\begin{center}
\epsfxsize=1\hsize
\epsffile{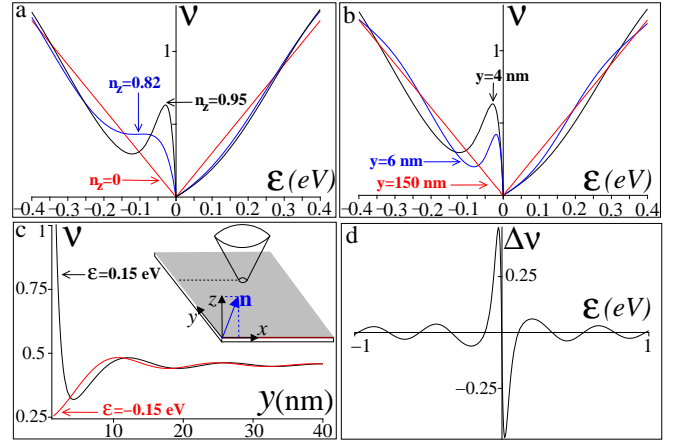}
\end{center}
\caption{(Color online) Local density of states in units of $3/4\pi$ eV$^{-1}$nm$^{-2}$:  
{\bf (a)} vs. energy for different $n_z$ at $y=4$ nm; 
{\bf (b)} vs. energy at different positions for $n_z=0.95$; 
{\bf (c)} vs. position for opposite-sign energies and $n_z=0.95$. 
Inset: local STS geometry and orientation of the unit vector ${\bf n}$ 
[Eq.~(\ref{K})] determining the boundary condition (\ref{BC}). 
{\bf (d)} Asymmetric local DOS 
$\Delta \nu\equiv\nu(\varepsilon)-\nu(-\varepsilon)$ 
vs. energy at $y=5$ nm for $n_z=0.95$. 
The data are for $v=10^6$ ms$^{-1}$ and $n_x<0$. 
}
\label{LDOS}
\end{figure}
%%%%%%%%%%%%%%%%%%%%%%%%%%%%%%%%%%%%%%%%%%%%%%%%%%%%%%%%%%%%%%%%%%%%%%%
%%%%%%%%%%%%%%%%%%%%%%%%%%%%%%%%%%%%%%%%%%%%%%%%%%%%%%%%%%%%%%%%%%%%%%%

\subsection{Energy and position dependence of the local DOS: Analysis}
\label{Asymmetry}

In Fig.~\ref{LDOS} we plot the local DOS~(\ref{nu_1}) 
as a function of energy $\varepsilon$ (in eV) and position $y$ (in nm)
for the Fermi velocity $v=10^6$ms$^{-1}$. 
These units and parameters are typical for STS in graphene.    
Panel~(a) shows an asymmetric peak, due to the edge states, 
emerging on top of the linear DOS as the boundary condition varies from 
the Berry-Mondragon type ($n_z=0$) to the zigzag type ($|n_z|\to 1$). 
In the latter case, the DOS~(\ref{nu_1}) still fails 
to recover the particle-hole symmetry because of the broken parity [see, Eq.~(\ref{Z})],
which is formally described by the singular energy-dependent factor $\Theta(\varepsilon n_xn_z)$.
The crossover between the Berry-Mondragon and zigzag cases  
can in principle be induced by a staggered mean-field sublattice potential 
whose strength is parametrized by 
the angle between ${\bf n}$ and ${\bf\hat z}$.~\cite{Akhmerov08}

For  $|n_z|\to 1$ the edge modes are much slower than the bulk ones, 
and have a small characteristic energy, $\approx\hbar v/2y\times |n_x/n_z|$. 
This can explain the observed DOS asymmetry on the scales of  $20-50$ meV~\cite{Niimi06}.
For $n_z=0.95$, the peak position, $\varepsilon\approx 25$ meV, 
and its overall behavior [panel~(b)] agree very well with 
the observations (see, e.g., Fig.~5 in Ref.~\onlinecite{Niimi06}).
As we neglect possible level broadening, 
the peak looks somewhat higher and narrower than in the experiment. 
Also, in agreement with the findings of Li et al.~\cite{Li08}, 
$\nu(\varepsilon)$ approaches the symmetric Dirac DOS away from the edge.  
The position dependence of the DOS 
[panel~(c)] shows that at the edge  
$\nu$ reaches either a maximum or a minimum 
depending on the presence or absence of the exponential term in 
Eq.~(\ref{nu_1}), which is controlled only by the sign of $\varepsilon$.
The Dirac waves, incident from the bulk and reflected from the edge, interfere 
yielding an oscillatory contribution (third term)  in the DOS~(\ref{nu_1}),   
decaying as $y^{-3/2}$ on the scale of 10-20 nm. 
Panel~(d) demonstrates similar oscillations in the energy dependence.  

%%%%%%%%%%%%%%%%%%%%%%%%%%%%%%%%%%%%%%%%%%%%%%%%%%%%%%
%%%%%%%%%%%%%% p %%%%%%%%%%%%%%%%%%%%%%%%%%%%%%%%%%%%%%%
%%%%%%%%%%%%%%%%%%%%%%%%%%%%%%%%%%%%%%%%%%%%%%%%%%%%%%
\begin{figure}[t]
\begin{center}
\epsfxsize=1\hsize
\epsffile{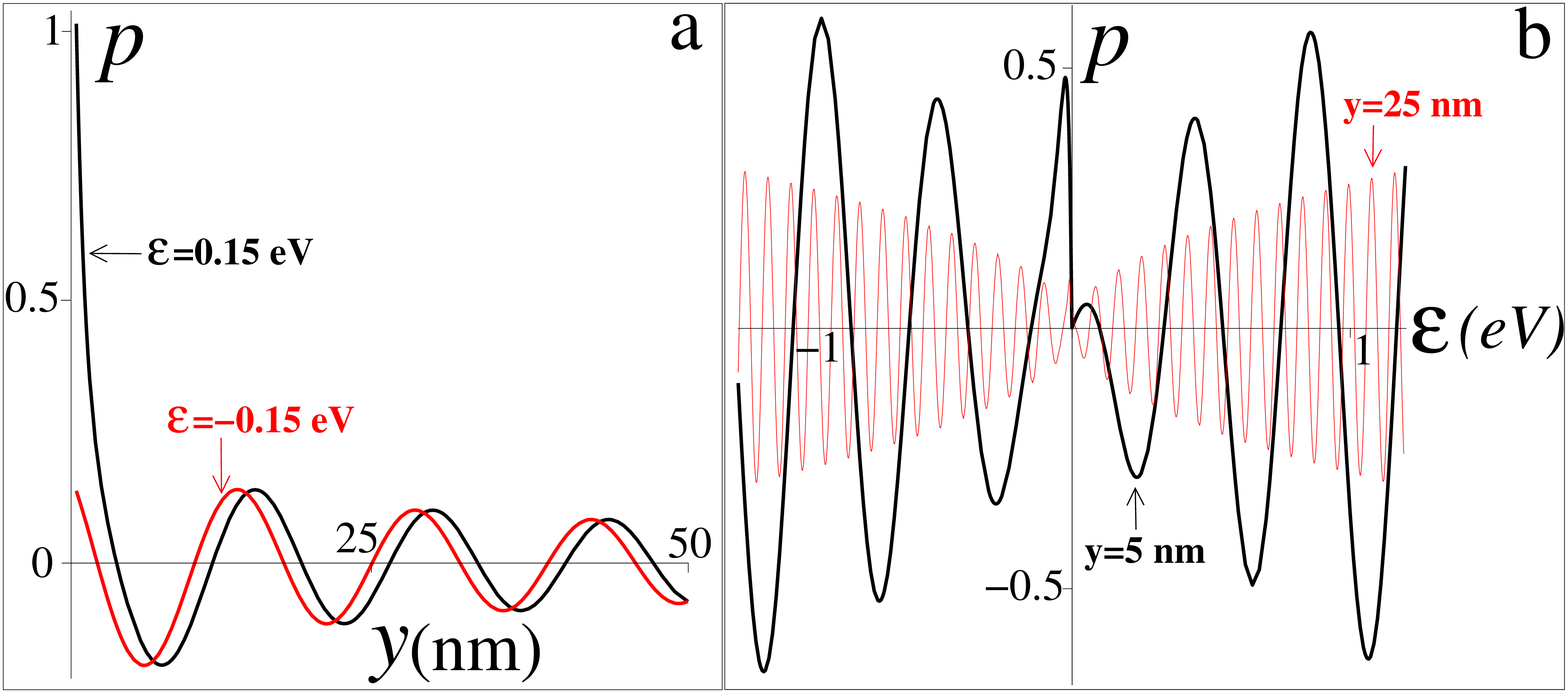}
\end{center}
\caption{(Color online)  
Chiral pseudospin polarization
{\bf (a)} vs. position   
and 
{\bf (b)} vs. energy for $n_z=0.95$ and $n_x<0$. 
}
\label{pFig}
\end{figure}
%%%%%%%%%%%%%%%%%%%%%%%%%%%%%%%%%%%%%%%%%%%%%%%%%%%%%%
%%%%%%%%%%%%%%%%%%%%%%%%%%%%%%%%%%%%%%%%%%%%%%%%%%%%%%

\subsection{ Local chiral pseudospin polarization }
\label{CPP}

To conclude the analysis of the local properties,  
in Fig.~\ref{pFig} we plot the CPP given by Eq.~(\ref{p_1}). 
From Fig.~\ref{pFig}(a) we see that the CPP has a purely boundary origin 
as it decays to zero in the bulk.  
Apart from the presence of the oscillations,  
both position and energy dependences of the local CPP 
differ significantly from the corresponding behaviors of the local DOS 
[cf. Figs.~\ref{pFig} and~\ref{LDOS}(c,d)].
Although not obvious in the local quantities 
$\nu(\varepsilon,y)$ and $p(\varepsilon,y)$, 
in the next section we establish a direct relation 
between the appropriately defined global CPP and DOS.

\section{ Global edge DOS and pseudospin polarization }
\label{Global}

\subsection{Relation between the edge DOS and pseudospin polarization}
\label{Relation}

Let us define the DOS and CPP of a finite region of space, $0\leq y \leq w$, 
as the following dimensionless integrals,  
\begin{eqnarray}
 &&
N_e(\varepsilon,w)=hv\int_0^w dy\, \left( \nu(\varepsilon,y) - \frac{ 2|\varepsilon| }{ \pi \hbar^2 v^2 } \right), 
\label{Ne_def}\\
&&
P(\varepsilon,w)=hv\int_0^w dy\, p(\varepsilon,y).
\label{P_def}
\end{eqnarray} 
In the first equation we subtract the bulk Dirac DOS, 
so that $N_e(\varepsilon,w)$ contains the contribution of the edge states only.  
By contrast to their  local counterparts, it is convenient to call 
$N_e(\varepsilon,w)$ and $P(\varepsilon,w)$ 
the global edge DOS and global CPP, respectively. 

Inserting Eqs.~(\ref{nu_1}) and  (\ref{p_1}) into Eqs.~(\ref{Ne_def}) and  (\ref{P_def}) 
and integrating over position $y$,  we find
\begin{eqnarray}
&&
N_e(\varepsilon)=
	 \frac{ 2\Theta\left( \varepsilon\, n_{x} n_{z} \right) 
        }{|n_{x}| }
        \left[
        1-{\rm e}^{ 
                        -\frac{ 2w }{ \hbar v } \left|\varepsilon \frac{ n_{z}  }{ n_{x}  } \right|  
                        }
       \right]
        -\frac{ 2 n_z{\rm sgn}\,\varepsilon }{ \pi }  
\label{Ne}\\
&&
        \times \int\limits_0^{ \frac{\pi}{2} } d\gamma 
\left[ 
           \frac{n_z}{ \sin\gamma}
              \frac{ 
              \sin\left( \frac{2\varepsilon w}{\hbar v } \sin\gamma \right)
	            }
	            { n^2_{z}+\tan^2\gamma }
         +  n_{x}
              \frac{
                1-\cos\left( \frac{2\varepsilon w}{\hbar v } \sin\gamma \right)
	            }
	            { n^2_{z}+\tan^2\gamma }
\right],
\nonumber	
\end{eqnarray}
\begin{eqnarray}
&&
P(\varepsilon)=
	 \frac{ 2n_z\Theta\left( \varepsilon\, n_{x} n_{z} \right) 
                }{|n_{x}| }
                \left[
                1-{\rm e}^{ 
                                 -\frac{ 2w }{ \hbar v } \left|\varepsilon \frac{ n_{z}  }{ n_{x}  } \right|  
                                }
                \right]
        + \frac{ 2{\rm sgn}\,\varepsilon }{ \pi } 
\label{P}\\
&&
        \times \int\limits_0^{ \frac{\pi}{2} } d\gamma 
\left[     
           \frac{n_z}{ \sin\gamma}
              \frac{ 
              \sin\left( \frac{2\varepsilon w}{\hbar v } \sin\gamma \right)
	            }
	            { 1 + n^2_{z}\cot^2\gamma }
        + n_{x}
            \frac{
                1-\cos\left( \frac{2\varepsilon w}{\hbar v } \sin\gamma \right)
	            }
	            { 1 + n^2_{z}\cot^2\gamma }
\right].
\nonumber	
\end{eqnarray}
It is instructive to discuss first the limit $w\to\infty$, when the integrals in Eqs.~(\ref{Ne}) 
and (\ref{P}) can be evaluated analytically.  
In this case the integrals with the rapidly oscillating cosine function vanish, 
while those containing the sine function should be evaluated with care since 
the ratio $\sin\left( \frac{2\varepsilon w}{\hbar v } \sin\gamma \right)/\sin\gamma $ 
becomes singular, $\pi\, {\rm sgn}(\varepsilon)\, \delta( \sin\gamma )$ as $w\to\infty$.
After integrating with the delta function $\delta( \sin\gamma )$, we have
\begin{eqnarray}
&&
N_e(\varepsilon) =
	 \frac{ 2\Theta\left( \varepsilon\, n_{x} n_{z} \right) 
        }{|n_{x}| }
        -1 
        - \frac{ 2 n_x n_z {\rm sgn}\,\varepsilon }{ \pi }
          \int\limits_0^{ \frac{\pi}{2} } 
                \frac{
                                       d\gamma 
	               }
	              { n^2_{z}+\tan^2\gamma },
\nonumber\\
&&
P(\varepsilon) =
	 \frac{ 2n_z\Theta\left( \varepsilon\, n_{x} n_{z} \right) 
        }{|n_{x}| }  
        + \frac{ 2 n_x {\rm sgn}\,\varepsilon }{ \pi }
            \int\limits_0^{ \frac{\pi}{2} } 
                 \frac{
                                       d\gamma 
	               }
	              { 1 + n^2_{z}\cot^2\gamma } .
\nonumber
\end{eqnarray}
The remaining integrals are easy to evaluate~\cite{Integrals}. 
The results are 
\begin{eqnarray}
N_e(\varepsilon)=\frac{ 1 + n_z {\rm sgn}(\varepsilon\, n_x)  }{ |n_x| } -1,
\label{Ne_1}
\end{eqnarray}
\begin{eqnarray}
P(\varepsilon)=\frac{ {\rm sgn}(\varepsilon\, n_x) + n_z}{|n_x|},  
|P(\varepsilon)|=\frac{ 1 + n_z {\rm sgn}(\varepsilon\, n_x) }{|n_x|}.
\label{P_1}
\end{eqnarray}
From Eqs.~(\ref{Ne_1}) and (\ref{P_1}) we find the relation between $N_e$ and $P$, 
\begin{eqnarray}
N_e(\varepsilon)=| P(\varepsilon) | - 1.
\label{DOS}
\end{eqnarray}
Being a scalar, $N_e$ does not depend on the sign of $P$ which is
reversed under the transformation ${\bf n}\to -{\bf n}$. 

%%%%%%%%%%%%%%%%%%%%%%%%%%%%%%%%%%%%%%%%%%%%%%%%%%%%%%%%%%%%%%%%%%%%%
%%%%%%%%%%%%%% Corr %%%%%%%%%%%%%%%%%%%%%%%%%%%%%%%%%%%%%%%%%%%%%%
%%%%%%%%%%%%%%%%%%%%%%%%%%%%%%%%%%%%%%%%%%%%%%%%%%%%%%%%%%%%%%%%%%%%%
\begin{figure}[t]
\begin{center}
\epsfxsize=0.7\hsize
\epsffile{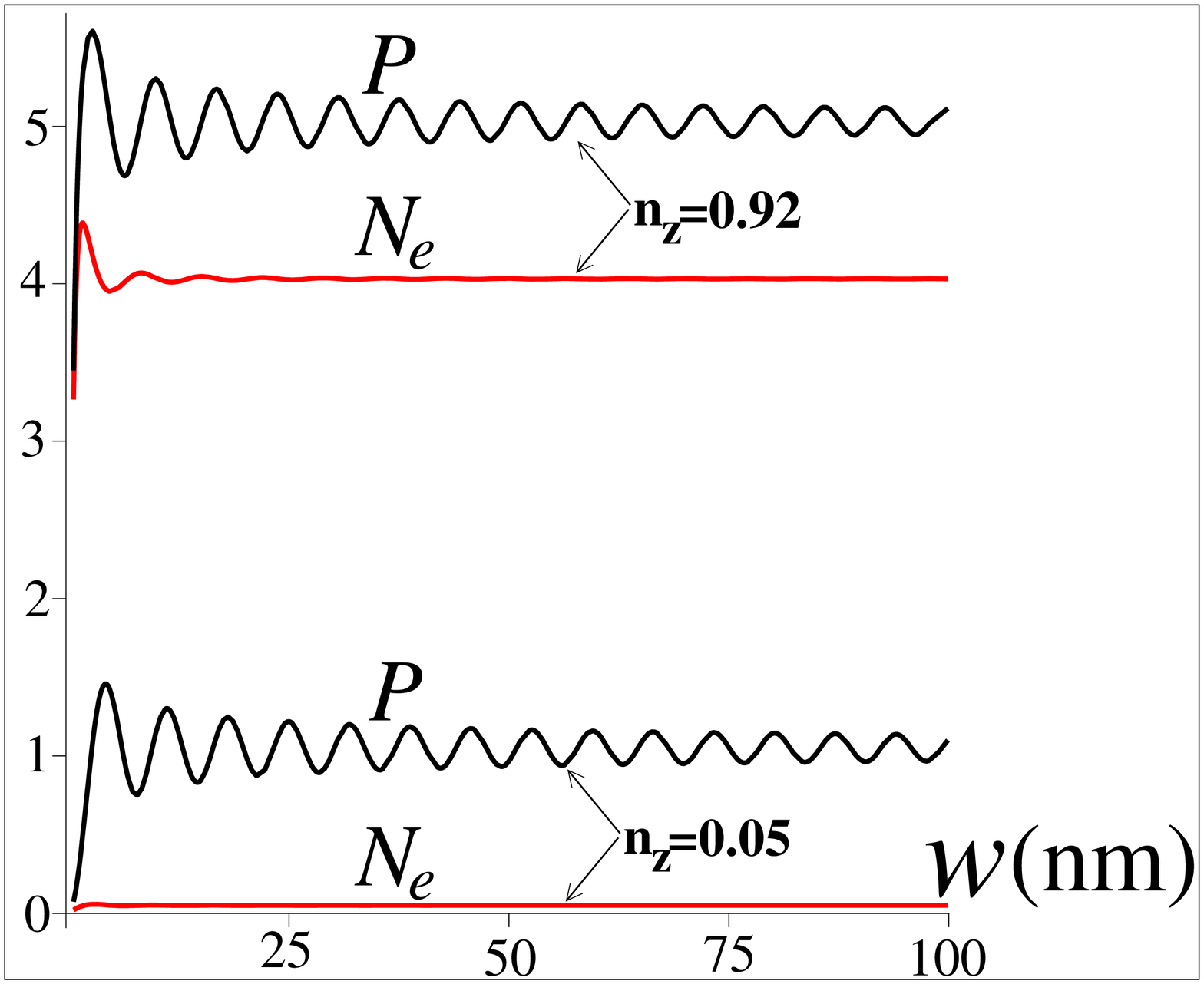}
\end{center} 
\caption{(Color online)
Typical behavior of the global edge DOS, $N_e$ and 
chiral pseudospin polarization,  $P$ as functions of the width, $w$ 
(see, Eqs.~(\ref{Ne_def}) -- (\ref{P}) ).
}
\label{Corr}
\end{figure}
%%%%%%%%%%%%%%%%%%%%%%%%%%%%%%%%%%%%%%%%%%%%%%%%%%%%%%%%%%%%%%%%%%%%%%%
%%%%%%%%%%%%%%%%%%%%%%%%%%%%%%%%%%%%%%%%%%%%%%%%%%%%%%%%%%%%%%%%%%%%%%%

According to Eq.~(\ref{DOS}), for the zigzag graphene edge $|n_z|\to 1 (n_x\to 0)$ 
the absolute value of the CPP becomes equal to the edge DOS:
\begin{eqnarray}
N_e(\varepsilon)\approx | P(\varepsilon) |\approx \frac{ 2\Theta( \varepsilon n_xn_z)  }{ |n_x| }.
\label{Ne_2}
\end{eqnarray} 
This means that the CPP can, in principle, be detected through measurements of the global edge DOS. 
The latter, in turn, can be probed by tunneling, as we discuss in the next subsection.
Before going to that question, we wish to point out that the correlation between $P$ and $N_e$ 
exists for finite values of $w$ as well. 
Figure~\ref{Corr} shows the functions  $N_e(w)$ [Eq.~(\ref{Ne})] and  $P(w)$ [Eq.~(\ref{P})], 
clearly approaching the relation~(\ref{DOS}) for $w\geq 100$ nm. 
Note that the non-oscillatory components of $P$ and $N_e$ obey the relation~(\ref{DOS}) 
at much smaller $w$.

\subsection{Tunneling spectroscopy}
\label{Non}

Our proposal for tunneling spectroscopy of the global edge DOS exploits 
the particle-hole asymmetric nonanalytic energy dependence of $N_e(\varepsilon)$ 
[Eqs.~(\ref{Ne_1}) and (\ref{Ne_2})].  
It is essential that the particle-hole asymmetry persists in the case of zigzag-terminated graphene 
($|n_z|\to 1, n_x\to 0$) because this is an experimentally accessible system.   

\begin{figure}[t]
\begin{center}
\epsfxsize=1\hsize
\epsffile{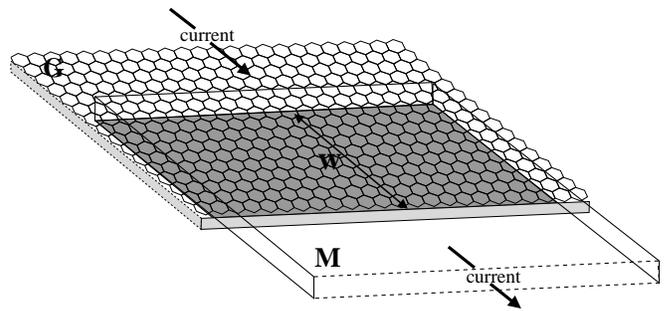}
\end{center} 
\caption{(Color online)
Suggested tunneling device for determining the density of the edge states in graphene.  
A metallic film (M) is deposited on top of a zigzag-terminated graphene sheet (G) 
forming a strip-like lateral contact of width, $w$.  
The device resistance is assumed to be determined by the tunnel barrier (dark gray area) 
so that the voltage drop $V$ predominantly occurs between the overlapping parts of M and G. 
}
\label{Tunnel}
\end{figure}

It is known~\cite{Giaever60,Mahan} that a strongly energy-dependent DOS reflects in 
the differential electric conductance, ${\rm g}(V)$ of a tunnel junction between the system of interest 
and a metal where the DOS is almost constant near the Fermi energy. 
Here we consider a lateral tunnel contact between a zigzag-terminated graphene sheet and 
a metallic film, as shown in Fig.~\ref{Tunnel}.  
It is assumed that the voltage drop, $V$, occurs predominantly across 
the tunnel barrier in the contact area, which determines the junction resistance.    
Under such condition, the conductance can be calculated using the tunneling Hamiltonian approach, 
which is well described in the literature (e.g. Ref.~\onlinecite{Mahan}),   
with the following result:
\begin{eqnarray}
&&
{\rm g}(V,T)={\rm g}_0 \int_{-\infty}^{\infty}
	d\varepsilon\, N(\varepsilon) \frac{\partial f(\varepsilon -eV,T)}{\partial (eV)},
\label{g}
\end{eqnarray} 
Here $N(\varepsilon)$ is the DOS of graphene in the contact region (dark gray area in Fig.~\ref{Tunnel}), 
$f(\varepsilon -eV,T)$ is the Fermi-Dirac distribution of the tunneling quasiparticles  
at voltage $V$ and temperature $T$, and the constant ${\rm g}_0$ absorbs the
energy-independent parameters of the metal and tunnel barrier. 
As we are interested in the low energy regime $V,T\to 0$, in Eq.~(\ref{g}) we can neglect 
inelastic tunneling processes (e.g. phonon emission).~\cite{Mahan}   

%%%%%%%%%%%%%%%%%%%%%%%%%%%%%%%%%%%%%%%%%%%%%%%%%%%%%%%%%%%%%%%%%%%%%
%%%%%%%%%%%%%% Tunnel %%%%%%%%%%%%%%%%%%%%%%%%%%%%%%%%%%%%%%%%%%%%%%
%%%%%%%%%%%%%%%%%%%%%%%%%%%%%%%%%%%%%%%%%%%%%%%%%%%%%%%%%%%%%%%%%%%%%
%%%%%%%%%%%%%%%%%%%%%%%%%%%%%%%%%%%%%%%%%%%%%%%%%%%%%%%%%%%%%%%%%%%%%%%
%%%%%%%%%%%%%%%%%%%%%%%%%%%%%%%%%%%%%%%%%%%%%%%%%%%%%%%%%%%%%%%%%%%%%%%

The DOS $N(\varepsilon)$ contains the contributions of both bulk and edge states. 
Since the bulk DOS is a symmetric function of energy ($\propto |\varepsilon|$), 
it can be eliminated  by taking the difference:
\begin{eqnarray}
&&
 \Delta {\rm g}(V,T) = {\rm g}(V,T)-{\rm g}(-V,T)= 
\nonumber\\
&&
={\rm g}_0\int_{-\infty}^{\infty}
	d\varepsilon\, [N_e(-\varepsilon)-N_e(\varepsilon)]\frac{\partial f(\varepsilon -eV,T)}{\partial\varepsilon}.
\label{Delta_g}
\end{eqnarray} 
It contains only the particle-hole asymmetic edge DOS, $N_e(\varepsilon)$, given by Eq.~(\ref{Ne}), 
where $w$ coincides with the width of the lateral tunnel contact [see, Fig.~\ref{Tunnel}]. 

\begin{figure}[t]
\begin{center}
\epsfxsize=0.6\hsize
\epsffile{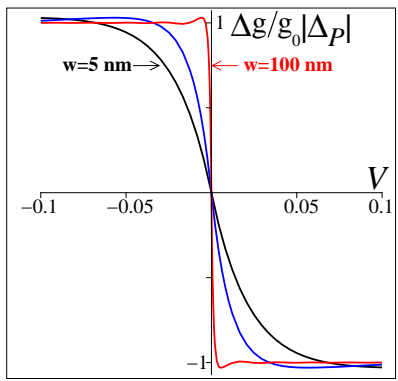}
\end{center} 
\caption{(Color online)
Zero-temperature conductance $\Delta {\rm g}={\rm g}(V)-{\rm g}(-V)$ vs. voltage (in volts) 
for different widths of the tunnel contact, $w$ [see, also Fig.~\ref{Tunnel}].
}
\label{Dg}
\end{figure}

In the limit $w\to\infty$, we use Eqs.~(\ref{P_1}) and ~(\ref{DOS}) to evaluate the integral in Eq.~(\ref{Delta_g}):   
\begin{eqnarray}
&&
\Delta {\rm g}(V,T)={\rm g}_0\, \Delta_P\,\tanh\frac{ eV }{ 2k_BT },
\label{Cond}\\
&&
\Delta_P=|P|_{\varepsilon>0}-|P|_{\varepsilon<0}=2\frac{n_z}{n_x}.
\label{Delta_P}
\end{eqnarray} 
The conductance asymmetry  $\Delta {\rm g}(V,T)$ reflects  
the nonequilibrium quasiparticle accumulation that builds up 
near the graphene edge in response to the current flow between the systems. 
For $|eV|>2k_BT$ the conductance $\Delta {\rm g}(V,T)$ saturates at $\pm {\rm g}_0\Delta_P$,
where $\Delta_P$ is the difference in the absolute values of the CPP for 
the positive- and negative-energy edge states. 
Such a nonlinear behavior can be used to detect the edge state as well as  
the existence of the pseudospin polarization. 
At zero temperature  $T=0$ the voltage dependence in Eq.~(\ref{Cond}) 
becomes singular ($\propto {\rm sgn}(eV)$). 
This is specific to the $w=\infty$ limit. 
As shown in Fig.~\ref{Dg}, the singularity is smeared due to finiteness of the contact width, $w$, 
so that  $\Delta {\rm g}(V,0)$ saturates at voltages larger than the value $\propto w^{-1}$. 
The data in Fig.~\ref{Dg} are obtained by numerical integration of
Eqs.~(\ref{Delta_g}) and (\ref{Ne}) at $T\to 0$.

%%%%%%%%%%%%%%%%%%%%%%%%%%%%%%%%%%%%%%%%%%%%%%%%%%%%%%%%%%%%%%%%%%%%%
%%%%%%%%%%%%%% Delta g %%%%%%%%%%%%%%%%%%%%%%%%%%%%%%%%%%%%%%%%%%%%%%
%%%%%%%%%%%%%%%%%%%%%%%%%%%%%%%%%%%%%%%%%%%%%%%%%%%%%%%%%%%%%%%%%%%%%
%%%%%%%%%%%%%%%%%%%%%%%%%%%%%%%%%%%%%%%%%%%%%%%%%%%%%%%%%%%%%%%%%%%%%%%
%%%%%%%%%%%%%%%%%%%%%%%%%%%%%%%%%%%%%%%%%%%%%%%%%%%%%%%%%%%%%%%%%%%%%%%

\section{Summary and discussion}
\label{Discussion}

We have considered the boundary problem for 2D Dirac fermions, Eq.~(\ref{Eq}) and (\ref{BC}),  
in which the time-reversal invariance is preserved at the expense of the Dirac fermion parity.
Using the Green's function solution, we have shown that the broken parity 
manifests itself in the density of the edge states and their pseudospin polarization, 
both exhibiting a nonanalytic particle-hole asymmetry. 
The zigzag graphene edge with its inherent structural asymmetry is an example of 
the realization of the Dirac fermion parity breaking.
Taking into account additionally the potential-energy sublattice asymmetry near the zigzag edge~\cite{Akhmerov08}, 
we obtain the local DOS consistent with the experimental data of Niimi et al.~\cite{Niimi06}
We have also established a direct correspondence between the 
pseudospin polarization and the density of the edge states, 
and suggested how to detect them in a tunneling experiment. 
The proposal relies on the broken particle-hole symmetry  
resulting in an asymmetric nonlinear contribution to the conductance, Eq.~(\ref{Cond}), 
in a tunnel junction between zigzag-terminated graphene and a metallic film [see, Fig.~\ref{Tunnel}]. 

It is interesting to discuss possible implementations of the strong nonlinearity of the conductance, Eq.~(\ref{Cond}). 
We suggest that it could be used  for detecting weak electric signals and their polarity.
The operation of such a device would exploit the two different states of the tunnel junction 
corresponding to the conductance values at positive and negative bias voltages [see, also, Fig.~\ref{Dg}]. 
Let us assume that the system is initially in one of these states. 
Then, under externally induced change in the bias voltage 
the system can switch into the state with the other (lower or higher) value of the conductance. 
For the zigzag graphene edge ($|n_z|\to 1, n_x\to 0$), 
the conductance difference, Eqs.~(\ref{Cond}) and (\ref{Delta_P}), is very significant and, therefore, should be detectable. 
The progress in characterization of graphene edges~\cite{Girit09} may eventually lead to more understanding 
of how the interfaces needed to test our finding can be fabricated.

We noticed that even though the conductance, Eq.~(\ref{Cond}) vanishes 
for the Berry-Mondragon confinement~\cite{Berry87} ($n_z\to 0$),
this case is still nontrivial, because the pseudospin polarization 
$P(\varepsilon)$ (\ref{DOS}) is not zero: 
$P(\varepsilon)={\rm sgn}(\varepsilon n_x)$. 
This is confirmed by the numerical data for $n_z=0.05$ in Fig.~\ref{Corr}.  
The nonvanishing  $P(\varepsilon)$ comes from the oscillatory (interference) term in Eq.~(\ref{p}), 
decaying as $y^{-1/2}$ on distances 50-100 nm [see, Fig.~\ref{pFig}(a)]. 
For a given chirality (valley), the long-range polarization implies violation of 
the ${\cal T}$ symmetry on mesoscopic scales, 
which may have some connection to recent studies of the level statistics in graphene quantum dots.~\cite{Ponomarenko08,Wurm09}
Also, such long-range polarization  may coexist with magnetic correlations predicted for 
zigzag graphene edges,~\cite{Fujita96,Son06,Wimmer08,Yazyev08} 
since they are expected to decay on much shorter distances.~\cite{Yazyev08}

We thank H. Baranger, F. Guinea, M. I. Katsnelson and A. D. Mirlin for discussions. 
The work was supported by the Emmy-Noether Programme (DFG).

%%%%%%%%%%%%%%%%%%%%%%%%%%%%%%%%%%%%%%%%%%%%%%%%%%%%%%%%%%%%%%%%%%%%%%%


\begin{thebibliography}{00}
\bibitem{Armchair}
Although this leaves out, e.g., armchair-type edges, 
the treatment is general enough to elucidate the origin of the parity violation.
\end{thebibliography}
\end{document}